\documentclass[amsmath,amssymb,aps,prl,superscriptaddress,twocolumn,notitlepage]{revtex4-1}

\usepackage{graphicx}
\usepackage{latexsym}
\usepackage{textcomp}
\usepackage{multirow,booktabs}
\usepackage{amsfonts,amsmath,amssymb}
\usepackage{url}
\usepackage{hyperref}
\hypersetup{colorlinks=false,pdfborder={0 0 0}}
\usepackage[utf8]{inputenc}
\usepackage[english]{babel}

\usepackage{graphicx,amsmath,amssymb,color,array}
\usepackage[normalem]{ulem}
\usepackage{multirow}
\usepackage{times}

\usepackage{amsfonts}
\usepackage{latexsym}
\usepackage{amsmath}
\usepackage{amssymb}
\usepackage{amsfonts}
\usepackage{amsthm}
\usepackage{mathrsfs}
\usepackage{natbib}
\usepackage{color,verbatim,graphics}
\usepackage{psfrag}
\DeclareMathAlphabet{\mathrsfs}{U}{rsfs}{m}{n}
\DeclareMathAlphabet{\mathpzc}{OT1}{pzc}{m}{it}
\DeclareMathAlphabet{\matheus}{U}{eus}{m}{n}
\DeclareMathAlphabet{\mathbbold}{U}{bbold}{m}{n}

\newcommand{\ba}{\begin{eqnarray}}
\newcommand{\be}{\begin{equation}}
\newcommand{\ee}{\end{equation}}
\newcommand{\beq}{\begin{equation}}
\newcommand{\eeq}{  \end{equation}}
\newcommand{\bea}{\begin{eqnarray}}
\newcommand{\eea}{  \end{eqnarray}}

\newcommand{\ea}{\end{eqnarray}}
\newcommand{\ban}{\begin{eqnarray*}}
\newcommand{\ean}{\end{eqnarray*}}

\newcommand{\tr}{\operatorname{tr}}

\newcommand{\ket}[1]{\left|#1\right\rangle}
\newcommand{\bra}[1]{\langle#1|}

\newcommand{\ketbra}[2]{|#1\rangle\langle#2|}

\newcommand{\eg}{{\it{e.g.}~}}
\newcommand{\ie}{{\it{i.e.}~}}

\newcommand{\rA}{\mathrm{A}}

\newcommand{\rB}{\mathrm{B}}
\newcommand{\rV}{\mathrm{V}}

\newcommand{\tel}{\mathrm{tel}}
\newcommand{\sep}{\mathrm{sep}}
\newcommand{\cl}{\mathrm{cl}}

\begin{document}

\title{All entangled states can demonstrate non-classical teleportation}

\author{Daniel Cavalcanti}
\affiliation{ICFO-Institut de Ciencies Fotoniques, The Barcelona Institute of Science and Technology, 08860 Castelldefels (Barcelona), Spain}
\email{daniel.cavalcanti@icfo.es}

\author{Paul Skrzypczyk}
\affiliation{H. H. Wills Physics Laboratory, University of Bristol, Tyndall Avenue, Bristol, BS8 1TL, United Kingdom}

\author{Ivan \v{S}upi\'{c}}
\affiliation{ICFO-Institut de Ciencies Fotoniques, The Barcelona Institute of Science and Technology, 08860 Castelldefels (Barcelona), Spain}
\email{daniel.cavalcanti@icfo.es}

\begin{abstract}
Quantum teleportation, the process by which Alice can transfer an unknown quantum state to Bob by using pre-shared entanglement and classical communication, is one of the cornerstones of quantum information. The standard benchmark for certifying quantum teleportation consists in surpassing the maximum average fidelity between the teleported and the target states that can be achieved classically. According to this figure of merit, not all entangled states are useful for teleportation. Here we propose a new benchmark that uses the full information available in a teleportation experiment and prove that all entangled states can implement a quantum channel which can not be reproduced classically. We introduce the idea of non-classical teleportation witness to certify if a teleportation experiment is genuinely quantum and discuss how to quantify this phenomenon. Our work provides new techniques for studying teleportation that can be immediately applied to certify the quality of quantum technologies.
\end{abstract}

\maketitle
Quantum teleportation \cite{teleportation} is a cornerstone of quantum information science, and serves as a primitive in several quantum information tasks \cite{Swapping,repeaters,computation}. Since the first demonstrations \cite{Boschi98,Bownmeester97,Furusawa98}, quantum teleportation has been implemented in a variety of physical systems and has become a testbed for quantum information platforms \cite{TelepReview}.
In the ideal setting, quantum teleportation refers to the situation where Alice shares a maximally entangled state with Bob, which she uses, in combination with classical communication, to faithfully transmit a quantum state to Bob, even if that state is unknown to her.

In order to test that Alice and Bob are performing quantum teleportation, a third party, which we refer to as the verifier, provides quantum systems to Alice in states $\ket{\omega_x}$, $x = 1,\ldots, |x|$, which are unknown to her, and asks her to transmit these states to Bob. By applying a Bell-state measurement on the input system and her share of the maximally entangled state, Alice projects Bob's system into the states $\rho^\rB_{a|\omega_x}=U_a \ketbra{\omega_x}{\omega_x} U_a^{\dagger}$, where $U_a$ is a known unitary operation that depends on the outcome $a$ of the Bell state measurement. By classically communicating the outcome $a$ to Bob, he can correct the unwanted unitary $U_a$, and then send his system to the verifier, who is able to check whether it is the same as the one provided to Alice. Note that for the purpose of verification, it is completely equivalent if Alice communicates the outcome $a$ to the verifier instead of Bob, who can then check if the state Bob sent -- which is now uncorrected -- is equal to the state given to Alice, modulo the correction.

In any realistic teleportation scheme, the states and measurements used will not be perfect. In this case the states that Bob receives after Alice applies a measurement with POVM elements $M_a^{\rV\rA}$ on systems V and A are given by
\be\label{assemblage}
\rho_{a|\omega_x}^\rB=\frac{\tr_{\rV\rA} [(M_{a}^{\rV\rA}\otimes \openone^\rB)\cdot(\ketbra{\omega_x}{\omega_x}^\rV\otimes\rho^{\rA\rB})]}{p(a|\omega_x)},
\ee
where $\rho^{\rA\rB}$ is the state shared by Alice and Bob, and $p(a|\omega_x)=\tr [(M_{a}^{\rV\rA}\otimes \openone^\rB)\cdot(\ketbra{\omega_x}{\omega_x}^\rV\otimes\rho^{\rA\rB})]$ is the probability of the particular outcome $a$ given that the verifier gives to Alice the state $\ket{\omega_x}$.
\begin{figure}[t]
\includegraphics[width=0.8\columnwidth]{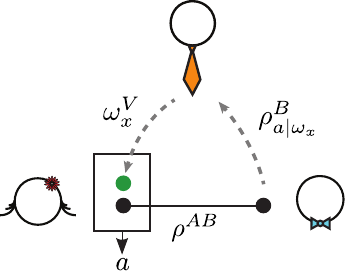}
\caption{(Color online). Teleportation scenario: Alice and Bob share a bipartite state $\rho^{\rA\rB}$. A verifier, who wants to check whether this state is entangled, sends systems in one of the states $\psi^\rV_x$ to Alice, and asks her to transmit it to Bob. Alice applies a global measurement on the state given to her by the verifier and her share of $\rho^{\rA\rB}$, which produces the states $\rho^\rB_{a|\omega_x}$ for Bob. The verifier has to determine if $\rho^{\rA\rB}$ is entangled based on the knowledge of  $\{\psi^\rV_x\}_x$ and $\{\rho^\rB_{a|\rho_x}\}_{a,x}$.}
\label{fig:teleportation}
\end{figure}
The standard figure-of-merit used to quantify how well such an teleportation scheme performs is the average fidelity between the input and output states of the process \cite{NielsenChuang},
\ba\label{e:av fidelity}
\overline{F}_{\tel}=\frac{1}{|x|}\sum_{a,x}p(a|\omega_x)\bra{\omega_x} U_a \rho^B_{a|\omega_x} U_a^{\dagger} \ket{\omega_x}.
\ea
Clearly, in the case of a perfect teleportation scheme, $\overline{F}_{\tel}=1$, while in real experiments one always obtains a smaller value. In the other direction, in a classical teleportation scheme -- one where Alice and Bob do not share any entanglement -- the maximum fidelity of teleportation that they can obtain, called the classical average fidelity, is denoted by $\overline{F}_{\cl}$. Thus, an imperfect teleportation scheme is certified to be non-classical if $\overline{F}_{\tel}>\overline{F}_{\cl}$ \cite{TelepReview}. It turns out that some entangled states can never lead to $\overline{F}_{\tel}>\overline{F}_{\cl}$, a famous example being bound entangled states \cite{Horodecki99}. Thus, according to this benchmark, these entangled states are useless for teleportation (although they can help in improving $\overline{F}_{\tel}$ of a combined state \cite{Masanes}).

However, notice that one has more information in a teleportation experiment than simply the value of $\overline{F}_{\tel}$. In particular, the verifier has access to $\{\ket{\omega_x}\}_x$, $\{\rho^\rB_{a|\omega_x}\}_{a,x}$ and $\{p(a|\omega_x)\}_{a,x}$. Whenever this data cannot be explained by a classical teleportation scheme then non-classical teleportation has clearly taken place. In principle, there could even exist a situation where $\overline{F}_{\tel}\leq \overline{F}_{\cl}$, but for which no classical teleportation scheme can explain the full data observed in the experiment.

The goal of this paper is twofold. First we propose a method to quantify the non-classicality of a teleportation scheme that uses the full data available. This method can be implemented by semi-definite programming (SDP) and provides non-classical teleportation witnesses, which generalise the average fidelity of teleportation. Second, we prove that every entangled state can be used to implement a teleportation scheme that is non-classical. This is true even with incomplete Bell state measurements, or when utilising inefficient detectors.

\textit{Quantifying non-classicality of teleportation.---}
For convenience, in what follows we will work with the set of unnormalised teleported states
\ba\label{eq: teleportation channel}
\sigma_{a|\omega_x}^\rB
&=&\tr_{\rV\rA} [(M_{a}^{\rV\rA}\otimes \openone^\rB)\cdot(\omega_x^\rV\otimes\rho^{\rA\rB})]\nonumber\\
&=&\tr_{\rV} [M_{a}^{\rV\rB}(\omega_x^\rV\otimes\openone^{\rB})],
\ea
where the state given to Alice by the verifier is now simply denoted $\omega_x^\rV$, which need not be a pure state, and
\be\label{eq: channel operators}
M_{a}^{\rV\rB}=\tr_\rA[(M_{a}^{\rV\rA}\otimes \openone^\rB)\cdot(\openone^\rV\otimes\rho^{\rA\rB})].
\ee
The normalisation factor $p(a|\omega_x)=\tr [\sigma_{a|\omega_x^\rV}]$ is the probability that Alice receives outcomes $a$ given that the input state was $\omega_x^\rV$. Equation \eqref{eq: teleportation channel} describes teleportation as a collection of channels from $\rV$ to $\rB$, labelled by $a$, that transform the input states $\omega_x^\rV$ into the (unnormalised) output states $\sigma_{a|\omega_x}^\rB$, according to the channel operators $M_{a}^{\rV\rB}$. Note that, due to the normalisation condition $\sum_a M_a^{\rV\rA} = \openone^{\rV\rA}$, the channel operators satisfy $\sum_a M_a^{\rV\rB} = \openone^\rV \otimes \rho^\rB$, where $\rho^\rB$ is Bob's reduced state, which can be seen as a no-signalling condition.

Consider now the case where $\rho^{\rA\rB}$ is a separable state, $\rho^{\rA\rB}=\sum_\lambda p_\lambda \rho^\rA_\lambda\otimes\rho^\rB_\lambda$, which we will see captures a completely general classical teleportation scheme. In this case the channel operators \eqref{eq: channel operators} become:
\ba\label{LHC}
M_{a}^{\rV\rB}&=&\sum_\lambda p_\lambda\tr_\rA[(M_{a}^{\rV\rA}\otimes \openone^\rB)\cdot(\openone^\rV\otimes \rho^\rA_\lambda\otimes\rho^\rB_\lambda)]\nonumber\\
&=&\sum_\lambda p_\lambda M_{a|\lambda}^{\rV}\otimes\rho^\rB_\lambda,
\ea
where $M_{a|\lambda}^{\rV}= \tr_\rA[ M_{a}^{\rV\rA} (\openone^\rV\otimes\rho^\rA_\lambda)]$. In this case Eq.~\eqref{eq: teleportation channel} becomes
\be
\sigma_{a|\omega_x}^\rB=\sum_\lambda p_\lambda \tr [M_{a|\lambda}^\rV\omega_x^\rV]\rho^\rB_\lambda.
\ee
This actually describes the most general classical teleportation scheme: a classical variable $\lambda$ is sampled from $p_\lambda$ and sent to Alice and Bob. Upon receiving $\lambda$ Alice measures the verifiers' system $\rV$ using the measurement operators $\{M_{a|\lambda}^\rV\}_{a}$ and obtains result $a$ according to the distribution $p(a|\omega_x^\rV,\lambda)= \tr [M_{a|\lambda}^\rV\omega_x^\rV]$. Bob, in turn, upon receiving $\lambda$ prepares the state $\rho^\rB_\lambda$ which he then sends to the verifier as the teleported state.

Given the structure of the this classical teleportation channel, we can test if a given set of teleportation data is nonclassical by solving the following optimisation problem:
\begin{align}
&\text{given} \quad\{\sigma_{a|\omega_x}^\rB\}_{a,x},\nonumber \\
&\mathcal{T}_R(\sigma_{a|\omega_x})=\min_{\substack{r,\{M_a^{\rV\rB}\}}} \quad r \label{e:telep robustness} \\
&\text{s.t.}\nonumber\\
&\frac{1}{(1+r)}\sigma_{a|\omega_x}^\rB + \frac{r}{(1+r)}\frac{1}{o_A}\frac{\openone^\rB}{d}=\tr_\rV [M_a^{\rV\rB} (\omega_x^\rV\otimes\openone^\rB)] \nonumber \\
&\hspace{6.5cm}\forall a,x,\nonumber \\
&M_a^{\rV\rB} \in \mathcal{S} \quad \forall a, \nonumber\\
&\sum_a M_a^{\rV\rB} = \openone^\rV \otimes \frac{\rho^\rB + r\frac{\openone^\rB}{d}}{1+r},\nonumber
\end{align}
where $o_A$ is the number of outcomes $a$, $\mathcal{S}$ denotes the set of separable operators (\ie of the form $\sum_\lambda \tau_\lambda \otimes \chi_\lambda$, with $\tau_\lambda \geq 0$ and $\chi_\lambda \geq 0$ for all $\lambda$). The optimal solution $r^*$ of this problem gives the minimum amount of `white noise' that has to be added to the teleportation data such that the mixture admits a classical scheme. We call $\mathcal{T}_R(\sigma_{a|\omega_x}) = r^*$ the random teleportation robustness of the data $\{\sigma_{a|\omega_x}^\rB\}_{a,x}$ \cite{footnote1,footnote2,footnote 3}.

Note that although the set of separable operators has a complicated structure \cite{EntReview}, we can nevertheless relax $\mathcal{S}$ in \eqref{e:telep robustness} to be the set of operators with positive partial transposition (PPT) \cite{Peres96}, which has a simple characterisation in terms of a single semidefinite constraint. In this case the above test becomes an instance of a strictly feasible semidefinite program \cite{VB94}, which can be easily solved with available software \cite{cvx}. Moreover, in the case of qubit teleportation, since the PPT criterion is necessary and sufficient for testing separability \cite{Horodecki96}, Eq.~\eqref{e:telep robustness} (without relaxation) is already an SDP, and therefore straightforward to solve. In higher dimensions other semidefinite relaxations of the set of separable operators have also been proposed and can be readily implemented \cite{DPS02}.

\textit{Every entangled state leads to non-classical teleportation.---}
As we show in \cite{supp}, in the case that (i) one of the Alice's measurement operators corresponds to a projection onto a maximally entangled state (\eg $B^{\rV\rA}_1=\ketbra{\Phi^+}{\Phi^+}$ with $\ket{\Phi^+}=\sum_{i=1}^{d} \ket{ii}^{\rV\rA}/\sqrt{d}$) and (ii) the inputs $\omega_x$ are tomographically complete, $\mathcal{T}_R(\sigma_{a|\omega_x})$ is proportional to $E_R(\rho^{\rA\rB})$ \cite{Robustness}, the random robustness of the state $\rho^{\rA\rB}$, defined as
\begin{align}
E_R(\rho^{\rA\rB}) = \min_{r,\sigma_S}& \quad r \label{e:ran ent robustness} \\
\text{s.t.}&\quad  \frac{1}{1+r}\rho^{\rA\rB}+\frac{r}{1+r}\frac{\openone}{d^2}=\Sigma_S,\nonumber \\
&\quad \Sigma_S \in \mathcal{S}. \nonumber
\end{align}
Moreover, if Alice's measurement is a full Bell-state measurement, then $\mathcal{T}_R(\sigma_{a|\omega_x})=E_R(\rho^{\rA\rB})$.

Since $E_R(\rho^{\rA\rB})$ is non-null if and only if $\rho^{\rA\rB}$ is entangled \cite{Robustness}, this result shows that every entangled state can lead to a non-classical teleportation data. Second, since the only requirement is that one of the Alice's measurement operators is a projection onto a maximally entangled state, the demonstration of non-classical teleportation can be done with partial Bell state measurements. This is experimentally good, since some setups naturally use these type of measurements due to the impossibility of performing a complete Bell state measurement with linear optics \cite{LCS99} or the use of inefficient detectors. Finally, it gives a one-to-one correspondence between how fragile the entanglement of a state is and how well this state can be used as a non-classical teleportation channel.

In Ref. \cite{supp}, we also prove a quantitative relation between the robustness of teleportation and the average fidelity of teleportation. Namely, for any set of teleported states coming from $\rho^{\rA\rB}$ we have that
\be\label{e: bound rob}
\mathcal{T}_R(\sigma_{a|\omega_x})\geq \frac{\overline{F}_\mathrm{tel}(\sigma_{a|\omega_x}) - \overline{F}_\mathrm{cl}}{\overline{F}_\mathrm{cl} - 1/d}.
\ee
This bound makes it clear that $\mathcal{T}_R(\sigma_{a|\omega_x})$ is a stronger than $\overline{F}_\mathrm{tel}$  as quantifier of non-classical teleportation for any set $\{\sigma_{a|\omega_x}\}$, since $\overline{F}_\mathrm{tel}(\sigma_{a|\omega_x}) >\overline{F}_\mathrm{cl}$  implies that $\mathcal{T}_R(\sigma_{a|\omega_x})>0$ \cite{footnote4}. Moreover this bound can be tight: In the case of perfect teleportation using a maximally entangled state, a tomographically complete set of inputs, and a Bell state measurement, the left-hand-side becomes $\mathcal{T}_R(\sigma_{a|\omega_x})=E_R(\rho^{\rA\rB})$ as discussed before. Moreover, since the state is maximally entangled we have that $E_R(\rho^{\rA\rB})=d$ \cite{Robustness}. The right-hand-side also equals $d$, since $\overline{F}_\mathrm{tel}=1$ and $\overline{F}_\mathrm{cl}=2/(d+1)$.

\textit{Non-classical teleportation witnesses.---}
An advantage of having a SDP formulation for teleportation is that it also provides linear constraints satisfied by any teleportation data that admits a classical scheme, which generalise the average fidelity of teleportation. These constraints work as \emph{non-classical teleportation witnesses}, which, similarly to the idea entanglement witnesses \cite{EntReview}, can be used to test the non-classicality of any experimental teleportation data. In Ref. \cite{supp} we show that the the random teleportation robustness $\mathcal{T}_R(\sigma_{a|\omega_x})$, given by \eqref{e:telep robustness} , has the following dual formulation \cite{VB94}:
 \begin{align} \label{e: dual}
	\text{given}& \quad\{\sigma_{a|\omega_x}^\rB\}_{a,x},\nonumber \\
	\max_{\{F_{a|\omega_x^\rB}\}, G^{\rV\rB}} &\quad \tr\sum_{a,x} F_{a|\omega_x}^\rB \sigma_{a|\omega_x}^\rB - \tr[G^\rB \rho^\rB] \nonumber \\
	\text{s.t.}&\quad 1 + \frac{1}{o_\rA d}\tr\sum_{a,x}F_{a|\omega_x}^\rB-\frac{1}{d}\tr G^{\rV\rB}  \geq 0, \\
	&\quad -\sum_x \omega_x^\rV \otimes F_{a|\omega_x}^\rB + G^{\rV\rB} \in \mathcal{W}\quad \forall a, \nonumber
\end{align}
The first constraint is a normalisation condition, while the second says that $W_a = -\sum_x \omega_x^\rV \otimes F_{a|\omega_x}^\rB + G^{\rV\rB}$ is an entanglement witness for all $a$.

See \cite{supp} for explicit examples of nonclassical teleportation witnesses.

\textit{Examples.---}
\begin{figure}[t]
\includegraphics[width=\columnwidth]{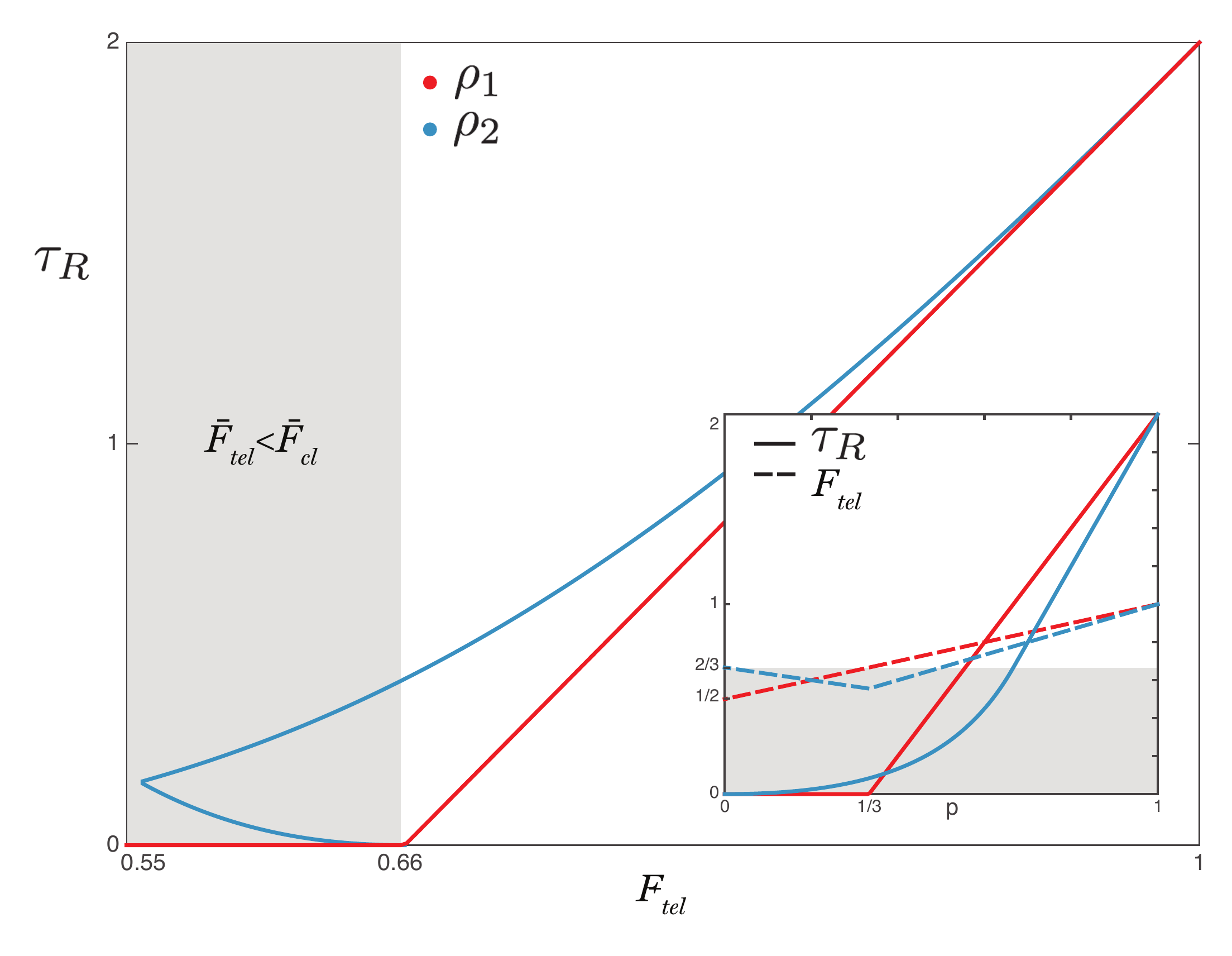}
\caption{(Color online). Average fidelity of teleportation \cite{Grudka} versus random teleportation robustness for $\rho^{\rA\rB}_1$ (red) - Eq. \eqref{e:werner} - and $\rho^{\rA\rB}_2$ (blue) - Eq. \eqref{e:AD}. The grey regions are shows the region where the average fidelity of teleportation is below the classical average fidelity $\bar{F}_{cl}=2/3$ \cite{MassarPopescu95}. The inset shows the same quantities as a function of the noise parameter $p$ for the two states.}
\label{f: examples}
\end{figure}
Let us discuss the relevance of the present results through two concrete examples. We consider teleportation of the states $ \{(\ket{0}\pm\ket{1})/\sqrt{2},(\ket{0}\pm i\ket{1})/\sqrt{2},\ket{0},\ket{1}\}$ (which form a state two-design) using the shared states
\be\label{e:werner}
\rho^{\rA\rB}_1=p\ketbra{\Phi^+}{\Phi^+}+(1-p)\frac{\openone^{\rA\rB}}{4}
\ee
and
\be\label{e:AD}
\rho^{\rA\rB}_{2}=p\ketbra{\Phi^+}{\Phi^+}+(1-p)\ketbra{01}{01}
\ee
and a full Bell state measurement. The results of the SDP \eqref{e:telep robustness} are provided in Fig \eqref{f: examples} as a function of the average fidelity of teleportation \eqref{e:av fidelity} when we vary $0\leq p\leq 1$. First notice that for the same values of $\overline{F}_{tel}$ the two states give different values for $\mathcal{T}_R$. This means that, although the two states perform equally as quantified by the average fidelity, when quantified instead by the random teleportation robustness $\rho^{\rA\rB}_{2}$ produces teleportation data which is more non-classical than $\rho^{\rA\rB}_{1}$ does. Second, there is a parameter region for which $\rho^{\rA\rB}_{2}$ is useless for teleportation according to $\overline{F}_{tel}$ (\ie $\overline{F}_{tel}\leq \overline{F}_{cl}$), but $\mathcal{T}_R$ still certifies that the teleportation data it produces could not arise from any classical teleportation scheme.

\textit{Connection to other notions of nonlocality}.---The present study also makes clear some connections between quantum teleportation and other ideas discussed in quantum foundations, such as EPR steering \cite{Steering,Steering Review} and Bell inequalities with quantum inputs \cite{Buscemi}. EPR steering is sometimes phrased in terms of a task where Bob wants to certify that he shares entanglement with Alice, but he does not trust her. He then asks her to perform some measurements on her share of the state and applies a test based on the post-measured  states he obtains. Notice that this is exactly the teleportation scenario as presented in Fig. \ref{fig:teleportation}, but with the crucial distinction that the inputs to Alice's measuring devices are classical variables $x$, as opposed to the quantum variables $\omega_x$ in teleportation. Crucially, due to this difference, not every entangled state is useful for demonstrating steering \cite{MT}. Another similar situation is the recently introduced Bell inequalities with quantum inputs \cite{Buscemi} (see also \cite{CHW13} for variations), which was later interpreted as the task of measurement-device-independent entanglement detection \cite{MDI-EW}. The scenario is the same as in quantum teleportation, but now Bob also applies a measurement with a quantum input to his share of the state. We thus see that teleportion relates to Bell inequalities with quantum inputs in exactly the same way that EPR steering relates to Bell nonlocality.

\textit{Conclusions.}--- In this letter we have studied quantum teleportation using the full data available in an experiment. We have shown that this allows us to test directly whether the data has any classical explanation via the method of semidefinite programming. Using the full data, every entangled state can be certified to implement non-classical teleportation, and we show that this can be tested in an experimentally friendly way using a teleportation witness. This overthrows the popular belief that not all entangled states are useful for teleportation (in particular bound entangled states), a conclusion which was based upon a single figure of merit, the average fidelity of teleportation, which our teleportation witnesses generalise.

\section{acknowledgements}
We thank A. Ac\'in, S. Popescu and J. Kolodynski for enlightening discussions, and F. Sciarrino for useful discussions on the experimental impact of our work. We also thank Kaifeng Bu for pointing out a mistake in a previous version. This work was supported by the Ram\'on y Cajal fellowship, Spanish MINECO (Severo Ochoa grant SEV-2015-0522 and FoQus), the AXA Chair in Quantum Information Science, the Generalitat de Catalunya (SGR875), Fundaci\'{o} Privada Cellex, ERC CoG QITBOX and the ERC AdG NLST.

\begin{widetext}

\section{Supplementary material}

\subsection{Every entangled state leads to non-classical teleportation}

Here we prove that when (i) Alice applies a partial Bell state measurement with measurement operators $M^{\rV\rA}_1=\ketbra{\Phi^+}{\Phi^+}$ and $M^{\rV\rA}_2=\openone-\ketbra{\Phi^+}{\Phi^+}$ and  (ii) the inputs $\omega_x$ are tomographically complete, then $\mathcal{T}_R(\sigma_{a|\omega_x})$ is proportional to $E_R(\rho^{\rA\rB})$. Then, the fact that the latter is non-zero for all entangled states, implies that the former is too, i.e. that every entangled state leads to non-classical teleportation.  For the sake of completeness let us redefine these quantities:
\begin{subequations}\label{e:telep robustness supp}
\begin{align}
&\text{given} \quad\{\sigma_{a|\omega_x}^\rB\}_{a,x},\nonumber \\
&\mathcal{T}_R(\sigma_{a|\omega_x})=\min_{\substack{r,\{M_a^{\rV\rB}\}}} \quad r  \\
&\text{s.t.}\nonumber\\
&\frac{1}{(1+r)}\sigma_{a|\omega_x}^\rB + \frac{r}{(1+r)}\frac{1}{o_A}\frac{\openone^\rB}{d}=\tr_\rV [M_a^{\rV\rB} (\omega_x^\rV\otimes\openone^\rB)] \quad\forall a,x, \label{e:telep robustness supp c1}\\
&M_a^{\rV\rB} \in \mathcal{S} \quad \forall a, \label{e:telep robustness supp c2}\\
&\sum_a M_a^{\rV\rB} = \openone^\rV \otimes \frac{\rho^\rB + r\frac{\openone^\rB}{d}}{1+r},\label{e:telep robustness supp c3}
\end{align}
\end{subequations}
where $o_A$ is the number of outcomes $a$, $\mathcal{S}$ denotes the set of separable operators (of the form $\sum_\lambda \tau_\lambda \otimes \chi_\lambda$, with $\tau_\lambda \geq 0$ and $\chi_\lambda \geq 0$ for all $\lambda$).

\begin{align}
&E_R(\rho^{\rA\rB}) = \min_{t,\Sigma_S} \quad t \label{e:ran ent robustness} \\
&\text{s.t.}\quad  \frac{1}{(1+t)}\rho^{\rA\rB}+\frac{t}{(1+t)}\frac{\openone}{d^2}=\Sigma_S,\\
&\Sigma_S \in \mathcal{S}. \nonumber
\end{align}

The main idea of the proof is to show that the optimisation problem \eqref{e:telep robustness supp} can be transformed into one which has the same form as \eqref{e:ran ent robustness} (with a parameter $t$ that is proportional to the parameter $r$ from \eqref{e:telep robustness supp}) if conditions (i) and (ii) are satisfied.

Starting with condition (i), focusing on the case where $a=1$, \ie $M^{\rV\rA}_1=\ketbra{\Phi^+}{\Phi^+}$, we can see that the left-hand-side of \eqref{e:telep robustness supp c1} can be written as
\begin{align}
&\frac{\sigma_{1|\omega_x}^\rB+r\frac{1}{o_A} \frac{\openone^\rB}{d}}{1+r}=\tr_{\rV\rA} \left[\ketbra{\Phi^+}{\Phi^+}^{\rV\rA}\otimes \openone^\rB \left(\omega_x^\rV\otimes\frac{\rho^{\rA\rB}+\frac{d^2r}{o_A}\frac{\openone^{\rA\rB}}{d^2}}{1+r}\right)\right]\\
&=\frac{1}{d}\tr_\rV\left[\left(\frac{\rho^{\rV\rB}+\frac{d^2 r}{o_A}\frac{\openone^{\rV\rB}}{d^2}}{1+r}\right)^{T_\rV}(\omega_x^\rV\otimes\openone^\rB)\right],
\end{align}
where $\rho^{\rV\rB}=\rho^{\rA\rB}$. In the last step we made use of the identity
\ba
\tr_\rA[(\ketbra{\Phi^+}{\Phi^+}^{\rV\rA}\otimes \openone^\rB)(\openone^\rV\otimes\tau^{\rA\rB})]=\frac{1}{d}[\tau^{\rV\rB}]^{T_\rV},
\ea
where $\tau^{\rV\rB} = \tau^{\rA\rB}$.

Thus, the first constraint  \eqref{e:telep robustness supp c1}  becomes
\ba
\frac{1}{d}\tr_\rV\left[\left(\frac{\rho^{\rV\rB}+\frac{d^2r}{o_A}\frac{\openone^{\rV\rB}}{d^2}}{1+r}\right)^{T_\rV}(\omega_x^\rV\otimes\openone^\rB)\right]=\tr_\rV [M_1^{\rV\rB} (\omega_x^\rV\otimes\openone^\rB)].
\ea
In the case that the set $\{\omega_x^\rV\}_x$ is a tomographically complete (condition (ii)) the only way that this equation can hold is if
\be\label{M1 def}
\frac{1}{d}\left(\frac{\rho^{\rV\rB}+\frac{d^2r}{o_A}\openone^{\rV\rB}/d^2}{1+r}\right)^{T_\rV}=M_1^{\rV\rB}.
\ee
This can equivalently be written as
\ba\label{e: Mtilde}
\frac{1}{1+\frac{d^2r}{o_A}}\rho^{\rV\rB}+\frac{\frac{d^2r}{o_A}}{1+\frac{d^2r}{o_A}}\frac{\openone^{\rV\rB}}{d^2}=\tilde{\sigma}^{\rV\rB},\nonumber\\
\ea
where $\tilde{\sigma}^{\rV\rB}\equiv d(1+r) (M_1^{\rV\rB})^{T_{\rV}}/(1+\frac{d^2r}{o_A})$. 

The constraint \eqref{e:telep robustness supp c2} implies that $\tilde{\sigma}^{\rV\rB}$ is a separable operator, since it is related to $M_1^{\rV\rB}$ by a rescaling and a partial transpose, both of which preserve separability.  

Moving on to $a=2$, where $M^{\rV\rA}_2=\openone-M^{\rV\rA}_1$, we see, similarly to above, that the left-hand-side of \eqref{e:telep robustness supp c1} can be written as
\begin{equation}
\frac{\sigma_{2|\omega_x}^\rB+r\frac{1}{o_A} \frac{\openone^\rB}{d}}{1+r}=\tr_\rV\left[\frac{1}{1+r}\left(\openone^\rV \otimes \rho^\rB-\frac{1}{d}\left(\rho^{\rV\rB}\right)^{T_\rV}+\frac{r}{o_A}\frac{\openone^{\rV\rB}}{d}\right)(\omega_x^\rV\otimes\openone^\rB)\right],
\end{equation}
from which it follows that
\begin{equation}\label{M2 def}
M_2^{\rV\rB} = \frac{1}{1+r}\left(\openone^\rV \otimes \rho^\rB-\frac{1}{d}\left(\rho^{\rV\rB}\right)^{T_\rV}+\frac{r}{o_A}\frac{\openone^{\rV\rB}}{d}\right).
\end{equation}
By direct substitution of $M_1^{\rV\rB}$ and $M_2^{\rV\rB}$ into the left hand side of constraint \eqref{e:telep robustness supp c3}, it is seen to be satified. It remains to show that $M_2^{\rV\rB}$ is a separable operator. To see this, first note that an alternative way to write $M_2^{\rV\rA}$ is 
\begin{equation}
M^{\rV\rA}_2=\sum_{i=1}^{d^2-1} (U_i^V \otimes\openone)^{-1}M^{\rV\rA}_1(U_i^V \otimes\openone),
\end{equation}
 where the $d^2-1$ unitary operators $U_i^V \otimes\openone$ generate all of the (mutually orthogonal) maximally entangled states that are orthogonal to $\ketbra{\Phi^+}{\Phi^+}$. Recalling the definition of $M_a^{\rV\rB}$, namely
\begin{equation}
M_{a}^{\rV\rB}=\tr_\rA[(M_{a}^{\rV\rA}\otimes \openone^\rB)\cdot(\openone^\rV\otimes\rho^{\rA\rB})]
\end{equation}
(see main text), this implies that  
\begin{equation}
M_{2}^{\rV\rB}=\sum_i (U_i^V \otimes\openone)^{-1}M^{\rV\rB}_1(U_i^V \otimes\openone)
\end{equation}
Therefore, if $M^{\rV\rB}_1$ is separable, so is $M^{\rV\rB}_2$, since each $(U_i^V \otimes\openone)^{-1}M^{\rV\rB}_1(U_i^V \otimes\openone)$ is separable (since local unitary transformations cannot generate entanglement), and $M^{\rV\rB}_2$ is a sum of such separable operators. 

To summarise, we have shown that if $M_1^{\rV\rB}$ as defined in \eqref{M1 def} is separable, then so too is $M_2^{\rV\rB}$ as defined in \eqref{M2 def}. Furthermore, with these definitions the no-signalling constraint \eqref{e:telep robustness supp c3} is satisfied. Finally, from \eqref{e: Mtilde}, we see that evaluating the teleportation robustness is equivalent to evaluating the entanglement random robustness of $\rho^{\rV\rB}$, with the specific relation
\begin{equation}
\mathcal{T}_R(\sigma_{a|\omega_x}) = \frac{o_A E_R(\rho^{\rA\rB})}{d^2}.
\end{equation} 
This proportionality between the teleportation random robustness and the entanglement random robustness implies that every entangled state demonstrates non-classical teleportation with a partial Bell state measurement and a tomographically complete set of states. In particular, every entangled state has non-zero entanglement random robustness, and thus the above shows that they also produce data which has non-zero teleportation random robustness, which completes the proof of the claim. 

We note finally that a similar analysis to above shows that if Alice instead performs a full Bell state measurement, with $d^2$ outcomes, then $\mathcal{T}_R(\sigma_{a|\omega_x}) = E_R(\rho^{\rA\rB})$.

\subsection{Non-classical Teleportation Witnesses}

We start noting that \eqref{e:telep robustness supp} is a convex optimisation problem, which gives the robustness to `white noise' of the teleportation data $\sigma_{a|\omega_x}$ before it admits a classical description, in terms of a separable effective POVM $M_a^{\rV\rB}$. This can be seen by noting that $\tilde{M}_a^{\rV\rB} = (1+r)M_a^{\rV\rB}$ is an unnormalised effective POVM that satisfies $\sum_a \tilde{M}_a^{\rV\rB} = \openone^\rV \otimes (\rho^\rB + r\frac{\openone^\rB}{d})$. It follows that an equivalent formulation of \eqref{e:telep robustness supp} is

\begin{align}
\text{given}& \quad\{\sigma_{a|\omega_x}^\rB\}_{a,x},\nonumber \\
\min_{r,\{\tilde{M}_a^{\rV\rB}\}}& \quad r \label{e:telep robustness final} \\
\text{s.t.}&\quad \sigma_{a|\omega_x}^\rB+r\frac{\openone^\rB}{o_\rA d}=\tr_\rV [\tilde{M}_a^{\rV\rB} (\omega_x^\rV\otimes\openone^\rB)] \quad\forall a,x,\nonumber \\
&\quad \sum_a \tilde{M}_a^{\rV\rB} = \openone^\rV \otimes (\rho^\rB + r\frac{\openone^\rB}{d})\nonumber \\
&\quad \tilde{M}_a^{\rV\rB} \in \mathcal{S} \quad \forall a, \nonumber \\
&\quad r \geq 0
\end{align}
which is now explicitly a convex optimisation problem in the variables $r$ and $\{M_a^{\rV\rB}\}_a$. The Lagrangian for this problem is
\begin{align}
\mathcal{L} &= r + \tr\sum_{a,x}F_{a|\omega_x}^\rB\left(\sigma_{a|\omega_x}^\rB + r \frac{\openone^\rB}{o_\rA d}- \tr_\rV [\tilde{M}_a^{\rV\rB} (\omega_x^\rV\otimes\openone^\rB)] \right) + \tr G^{\rV\rB}\left(\sum_a \tilde{M}_a^{\rV\rB} - \openone^\rV \otimes (\rho^\rB + r\frac{\openone^\rB}{d})\right) \nonumber \\
&\quad - \tr\sum_a H_a^{\rV\rB}\tilde{M}_a^{\rV\rB}- \mu r, \nonumber \\
	&= r\left(1 + \frac{1}{o_\rA d}\tr\sum_{a,x}F_{a|\omega_x}^\rB-\frac{1}{d}\tr G^{\rV\rB}  - \mu\right) + \tr \sum_a \tilde{M}_a^{\rV\rB}\left(-\sum_x \omega_x^\rB \otimes F_{a|\omega_x}^\rB  + G^{\rV\rB} -H_a^{\rV\rB}\right) \nonumber \\
	&\quad + \tr\sum_{a,x} F_{a|\omega_x}^\rB \sigma_{a|\omega_x}^\rB - \tr[G^\rB \rho^\rB] 
\end{align}
where $\{F_{a|\omega_x}^\rB\}_{a,x}$, $G^{\rV\rB}$, $\{H_a^{\rV\rB}\}_a$,  and $\mu$ are the Lagrange multipliers corresponding to each set of constraints respectively. By taking $H_a^{\rV\rB} \in \mathcal{W}$, where $\mathcal{W} = \{ W | \tr[W\rho^\mathrm{sep}] \geq 0, \forall \rho^\mathrm{sep} \in \mathcal{S}\}$ is the set of entanglement witnesses (operators which are positive on all separable operators), and $\mu \geq 0$, then by enforcing that the first and second brackets vanish, we can ensure $\mathcal{L} \leq r$ and thus the dual formulation of \eqref{e:telep robustness final} is
\begin{align}
	\text{given}& \quad\{\sigma_{a|\omega_x}^\rB\}_{a,x},\nonumber \\
	\max_{\{F_{a|\omega_x^\rB}\}, G^{\rV\rB}, \{H^{\rV\rB}_a\}} &\quad \tr\sum_{a,x} F_{a|\omega_x}^\rB \sigma_{a|\omega_x}^\rB - \tr[G^\rB \rho^\rB] \nonumber \\
	\text{s.t.}&\quad 1 + \frac{1}{o_\rA d}\tr\sum_{a,x}F_{a|\omega_x}^\rB-\frac{1}{d}\tr G^{\rV\rB}  - \mu = 0, \\
	&\quad -\sum_x \omega_x^\rV \otimes F_{a|\omega_x}^\rB  + G^{\rV\rB} -H_a^{\rV\rB} = 0\quad \forall a, \nonumber \\
	&\quad H_a^{\rV\rB} \in \mathcal{W} \quad \forall a, \nonumber \\
	&\quad \mu \geq 0. \nonumber 
\end{align}
It is seen that $\{H^{\rV\rB}_a\}_a$ and $\mu$ play the role of slack variables (they do not appear in the objective function), and can thus be eliminated from the problem, to arrive at the equivalent formulation
 \begin{align} \label{e:robustness dual}
	\text{given}& \quad\{\sigma_{a|\omega_x}^\rB\}_{a,x},\nonumber \\
	\max_{\{F_{a|\omega_x^\rB}\}, G^{\rV\rB}} &\quad \tr\sum_{a,x} F_{a|\omega_x}^\rB \sigma_{a|\omega_x}^\rB - \tr[G^\rB \rho^\rB] \nonumber \\
	\text{s.t.}&\quad 1 + \frac{1}{o_\rA d}\tr\sum_{a,x}F_{a|\omega_x}^\rB-\frac{1}{d}\tr G^{\rV\rB}  \geq 0, \\
	&\quad -\sum_x \omega_x^\rV \otimes F_{a|\omega_x}^\rB + G^{\rV\rB} \in \mathcal{W}\quad \forall a, \nonumber
\end{align}
By taking all dual variables to be proportional to the identity, it is straightforward to see that all constraints can be strictly satisfied, and hence strong duality holds. As such, the optimal value of the primal and dual formulations coincide. 

\subsection{Bounding the teleportation robustness by the average fidelity of teleportation}
In this section we show how to place a bound on the teleportation robustness based upon the average fidelity of teleportation. In order to do so we will need an alternative expression for the classical average fidelity of teleportation, which is a special instance of the more general question, that of finding the classical bound of a given teleportation witness. 

The classical bound of a teleportation witness (specified by the operators $\{F_{a|\omega_x}^\rB\}_{a,x}$ and $G^{\rV\rB}$) is the solution to the following optimisation problem
\begin{align}
f_\mathrm{cl} = \max_{\{M_a^{\rV\rB}\},\rho^\rB}& \quad \tr\sum_{a,x} M_a^{\rV\rB} (\omega_x^\rV \otimes F_{a|\omega_x}^\rB) - \tr[G^\rB \rho^\rB] \label{e:f classical} \\
\text{s.t}&\quad M_a^{\rV\rB} \in \mathcal{S} \quad \forall a, \nonumber \\
&\quad \sum_a M_a^{\rV\rB} = \openone^\rV \otimes \rho^\rB, \nonumber \\
&\quad \tr[\rho^\rB] = 1, \nonumber
\end{align}
which states the problem as an optimisation over the separable channel operators that arise when using a separable state in order to implement the most general classical teleportation scheme. 

By constructing the Lagrangian for the problem, as carried out explicitly above, one can show that the classical bound of teleportation has the following dual formulation
\begin{align}
f_\mathrm{cl} = \min_{\{X^{\rV\rB}\},\mu}& \quad \mu \label{e:f classical dual} \\
\text{s.t}&\quad - \sum_x \omega_x^\rV \otimes F_{a|\omega_x}^\rB + X^{\rV\rB}\in \mathcal{W} \quad \forall a, \nonumber \\
&\quad X^\rB - \mu \openone^\rB = G^\rB, \nonumber 
\end{align}
The primal formulation is seen to be strictly feasible, by choosing $M_a^{\rV\rB} = \frac{1}{o_\rA}\openone^\rV \otimes \openone^\rB/d$ and $\rho^\rB = \openone^\rB/d$, thus strong duality holds, and the value of the primal and dual formulations coincide. 

Specialising to the case of the average fidelity of teleportation, we have
\begin{align}
F_{a|\omega_x}^\rB &= \frac{1}{|x|} U_a^\dagger \omega_x^\rB U_a, & G^{\rV\rB} &= 0.
\end{align}

From \eqref{e:f classical dual}, we know that there exists $X^{*\rV\rB}$ such that $X^{*\rB} - \overline{F}_\mathrm{cl} \openone^\rB = 0$, from which is follows that $\tr X^{*\rV\rB} = d \overline{F}_\mathrm{cl}$, where $\overline{F}_\mathrm{cl}$ is the classical average fidelity of teleportation (which is nothing but the classical bound of the teleportation witness corresponding to the average fidelity of teleportation). 

We would now like to consider a new teleportation witness, which is based upon the witness for the average fidelity, of the following form
\begin{align}
\tilde{F}_{a|\omega_x}^\rB &= \alpha\frac{1}{|x|} U_a^\dagger \omega_x^\rB U_a, & \tilde{G}^{\rV\rB} &= \alpha X^{*\rV\rB}.
\end{align}
for $\alpha > 0$, that will be determined below. For this inequality, first note that 
\begin{align}\label{e:first constraint}
1 + \frac{1}{o_\rA d}\tr\sum_{a,x}\tilde{F}_{a|\omega_x}^\rB-\frac{1}{d}\tr \tilde{G}^{\rV\rB} &= 1 + \frac{\alpha}{o_\rA d}\tr\sum_{a,x}\frac{1}{|x|} U_a^\dagger \omega_x^\rB U_a -\frac{\alpha}{d}\tr  X^{*\rV\rB} \nonumber \\
&= 1 + \frac{\alpha}{d} -\alpha \overline{F}_\mathrm{cl} \nonumber \\
&= 0 \quad \text{if} \quad \alpha = \frac{1}{\overline{F}_\mathrm{cl} - 1/d},
\end{align}
which shows that the first constraint in \eqref{e:robustness dual} is satisfied (with equality) by this choice of $\{\tilde{F}_{a|\omega_x}^\rB\}_{a,x}$, $\tilde{G}^{\rV\rB}$ and $\alpha$. Note that $\alpha \geq 0$, since $\overline{F}_\mathrm{cl} \geq 1/d$, since the trivial classical strategy whereby Alice outputs $a$ with any probability distribution $P(a)$, and Bob always sends the maximally mixed state $\rho_{a|\omega_x}^\rB = \openone^\rB/d$ to the verifier, achieves fidelity of teleportation equal to $1/d$. 

Second, this inequality also satisfies
\begin{align}\label{e:second constraint}
-\sum_x \omega_x^\rV \otimes \tilde{F}_{a|\omega_x}^\rB + \tilde{G}^{\rV\rB} & = -\frac{\alpha}{|x|}\sum_x \omega_x^\rV \otimes U_a^\dagger \omega_x^\rB U_a + \alpha X^{*\rV\rB}, \nonumber \\
& = \alpha \left(-\frac{1}{|x|}\sum_x \omega_x^\rV \otimes U_a^\dagger \omega_x^\rB U_a +  X^{*\rV\rB}\right) \in \mathcal{W},
\end{align}
where the final inclusion in the set $\mathcal{W}$ follows from the fact that if $W_a \in \mathcal{W}$ then $\alpha W_a \in \mathcal{W}$ if $\alpha \geq 0$, and from \eqref{e:f classical dual} the term inside parentheses is guaranteed to be an entanglement witness for all $a$.

Altogether, \eqref{e:first constraint} and \eqref{e:second constraint} show that $\{\tilde{F}_{a|\omega_x}^\rB\}_{a,x}$, $\tilde{G}^{\rV\rB}$ (with the specific choice $\alpha = 1/(\overline{F}_\mathrm{cl} - 1/d)$) are a feasible set of dual variables for the problem \eqref{e:robustness dual}. Since they don't necessarily obtain the maximum, we therefore obtain the inequality
\begin{equation}
\mathcal{T}_R(\sigma_{a|\omega_x}) \geq \frac{\overline{F}_\mathrm{tel} - \overline{F}_\mathrm{cl}}{\overline{F}_\mathrm{cl} - 1/d}
\end{equation}
where $\overline{F}_\mathrm{tel} =\frac{1}{|x|}\sum_{a,x}\bra{\omega_x} U_a \sigma^B_{a|\omega_x} U_a^{\dagger} \ket{\omega_x}$ is the average fidelity of teleportation of the data. 

\subsection{Examples}

Consider teleportation of the states $\{\omega_x\}_x = \{\ket{0},\ket{1}, (\ket{0}\pm\ket{1})/\sqrt{2},(\ket{0}\pm i\ket{1})/\sqrt{2}\}$ using the two-qubit Werner state 
\be\label{e:werner}
\rho^{\rA\rB}=p\ketbra{\Phi^+}{\Phi^+}+(1-p)\frac{\openone^{\rA\rB}}{4}
\ee
and full Bell state measurement, $\{M_a^{\rV\rA}\}_a = \{\ketbra{\Phi^+}{\Phi^+},\ketbra{\Psi^+}{\Psi^+},\ketbra{\Phi^-}{\Phi^-},\ketbra{\Psi^-}{\Psi^-}\}$. The teleportation witness is given in Table~\ref{t:werner states}. We have $\tr\sum_{a,x}F_{a|\omega_x}^\rB\sigma_{a|\omega_x}^B=6(\tfrac{1}{3}-p)$, and therefore teleportation is certified for all $p>1/3$, which coincides with the separability bound of the state \eqref{e:werner}. Finally, we note that $\{W_a\}_a = \{4\ketbra{\Psi^-}{\Psi^-}, 4\ketbra{\Phi^-}{\Phi^-}, 4\ketbra{\Psi^+}{\Psi^+}, 4\ketbra{\Phi^+}{\Phi^+} \}$ and thus $\tr[W_a \rho_\sep] \geq 0$ as required by \eqref{e:robustness dual}.
\begin{table}[h]
$x$ \vspace{0.1cm}\\
$a$\hspace{0.1cm}\,
\begin{tabular}{c||c|c|c|c|c|c}
	$F_{a|\omega_x}^\rB$ & 0 & 1 & 2 & 3 & 4 & 5 \\ \hline\hline
0 & $\tfrac{\openone}{3}-X$ & $\tfrac{\openone}{3}+X$ & $\tfrac{\openone}{3}-Y$ & $\tfrac{\openone}{3}+Y$ & $\tfrac{\openone}{3}-Z$ & $\tfrac{\openone}{3}+Z$ \\ \hline
1 & $\tfrac{\openone}{3}-X$ & $\tfrac{\openone}{3}+X$ & $\tfrac{\openone}{3}+Y$ & $\tfrac{\openone}{3}-Y$ & $\tfrac{\openone}{3}+Z$ & $\tfrac{\openone}{3}-Z$ \\ \hline
2 & $\tfrac{\openone}{3}+X$ & $\tfrac{\openone}{3}-X$ & $\tfrac{\openone}{3}+Y$ & $\tfrac{\openone}{3}-Y$ & $\tfrac{\openone}{3}-Z$ & $\tfrac{\openone}{3}+Z$ \\ \hline
3 & $\tfrac{\openone}{3}+X$ & $\tfrac{\openone}{3}-X$ & $\tfrac{\openone}{3}-Y$ & $\tfrac{\openone}{3}+Y$ & $\tfrac{\openone}{3}+Z$ & $\tfrac{\openone}{3}-Z$
\end{tabular}
\caption{\label{t:werner states} Teleportation witness for the two-qubit Werner state \eqref{e:werner}. The verifier provides the states $\{\omega_x\}_x = \{\ket{0},\ket{1}, (\ket{0}\pm\ket{1})/\sqrt{2},(\ket{0}\pm i\ket{1})/\sqrt{2}\}$ to Alice. By measuring the observables $F_{a|\omega_x}^\rB$ when Bob forwards the state $\sigma^\rB_{a|\omega_x}$ to the verifier, the value $\tr\sum_{a,x}F_{a|\omega_x}^\rB\sigma_{a|\omega_x}^B=6(\tfrac{1}{3}-p)$ is obtained, which is negative for all $p > 1/3$. Thus all entangled two-qubit Werner states are witnessed as useful for teleportation.}
\end{table}

Consider now the so-called `tiles' bound entangled state \cite{UPB}:
\begin{equation}\label{e:tiles}
\rho_\mathrm{tiles}= \frac{1}{4}\left(\openone-\sum_{i=0}^{4}\ketbra{\phi_i}{\phi_i}\right),
\end{equation}
where the states $\ket{\phi_i}$ form an unextendible product basis (UPB): 
\begin{align}
 \ket{\phi_0} &= \frac{1}{\sqrt{2}}\ket{0}(\ket{0}-\ket{1}),& \ket{\phi_1} &= \frac{1}{\sqrt{2}}\ket{2}(\ket{1}-\ket{2}),\\
 \ket{\phi_2} &= \frac{1}{\sqrt{2}}(\ket{0}-\ket{1})\ket{2},& \ket{\phi_3} &= \frac{1}{\sqrt{2}}(\ket{1}-\ket{2})\ket{0},\nonumber
\end{align}
\vspace{-0.2cm}
\begin{equation*}
\ket{\phi_4} = \frac{1}{3}(\ket{0} + \ket{1} + \ket{2})(\ket{0} + \ket{1} + \ket{2}).
\end{equation*}
According to the benchmark based on the average fidelity this state is useless for teleportation \cite{Horodecki99}. 
For the set of input states $\{\omega_x\}_x = \{\ket{0}, \ket{2},(\ket{0}-\ket{1})/\sqrt{2},(\ket{1}-\ket{2})/\sqrt{2},(\ket{0} + \ket{1}+\ket{2})/\sqrt{3}, \openone/3\}$ and partial Bell state measurement $\{M_a^{\rV\rA}\}_a = \{\ketbra{\Phi^+}{\Phi^+}, \openone - \ketbra{\Phi^+}{\Phi^+}\}$ we generate the teleportation witness given in Table~\ref{t:BE state}. The state achieves the value $\tr\sum_{a,x}F_{a|\omega_x}^\rB \sigma_{a|\omega_x}^\rB = -\epsilon/3$, which shows that the bound entangled states are in fact useful for teleportation.

A couple of additional comments are in order. First, note that the set of input states $\{\omega_x\}_x$ in this instance is not even tomographically complete, and yet teleportation can nevertheless be certified. Second, here we considered only a partial Bell state measurement. Since $M_{1}^{\rV\rA}$ is a separable operator in this instance, it is for this reason that $F_{1|\omega_x}^\rB$ vanish. Finally, we note that $W_0 = \sum_x \omega_x^\rV \otimes F_{1|\omega_x}^\rB = \sum_i \ketbra{\phi_i}{\phi_i} - \epsilon \openone$ is precisely the entanglement witness which is violated by the `tiles' UPB state \eqref{e:tiles} for $\epsilon \in (0,0.02842]$ \cite{Guhne}. This demonstrates that the constraint in \eqref{e:robustness dual} is indeed satisfied.
\begin{table}
$x$ \vspace{0.1cm}\\
$a$\hspace{0.1cm}\,
\begin{tabular}{c||c|c|c|c|c|c}
$F_{a|\omega_x}^\rB$ & 0 & 1 & 2 & 3 & 4 & 5 \\ \hline\hline
0 & $\psi_2$ & $\psi_3$ & $\psi_1$ & $\psi_0$ & $\psi_4$ & $-3\epsilon\openone$ \\ \hline
1 & 0 & 0 & 0 & 0 & 0 & 0
\end{tabular}
\caption{\label{t:BE state} Teleportation witness for the two-qutrit bound entangled `tiles' state \eqref{e:tiles}. The verifier provides the states $\{\omega_x\}_x = \{\ketbra{0}{0}, \ketbra{2}{2},(\ket{0}-\ket{1})(\bra{0}-\bra{1})/2,(\ket{1}-\ket{2})(\bra{1}-\bra{2})/2,(\ket{0} + \ket{1}+\ket{2})(\bra{0} + \bra{1}+\bra{2})/3, \openone/3\}$ to Alice. Here, $\epsilon \in (0,0.02842]$. By measuring the observables $F_{a|\omega_x}^\rB$ when Bob forwards the state $\sigma^\rB_{a|\omega_x}$ to the verifier, the value $\tr\sum_{a,x}F_{a|\omega_x}^\rB\sigma_{a|\omega_x}^B=-\epsilon/3$ is obtained. This demonstrates the fact that, contrary to what is usually claimed, bound entangled states are useful for teleportation.}
\end{table}

\end{widetext}


\begin{thebibliography}{30}

\bibitem{teleportation}C. H. Bennett, G. Brassard, C. Cr\'epeau, R. Jozsa, A. Peres, and W. K. Wootters, \emph{Teleporting an unknown quantum state via dual classical and Einstein-Podolsky-Rosen channels}, Phys. Rev. Lett. {\bf70}, 1895 (1993).

\bibitem{Swapping}M. Zukowski, A. Zeilinger, M. A. Horne, and A. K. Ekert, \emph{``Event-ready-detectors'' Bell experiment via entanglement swapping}, Phys. Rev. Lett. {\bf71}, 4287 (1993).

\bibitem{repeaters}H.-J. Briegel, W. D\"{u}r, J. I. Cirac, and P. Zoller, \emph{Quantum repeaters: The role of imperfect local operations in quantum communication}, Phys. Rev. Lett. {\bf81}, 5932–5935 (1998).

\bibitem{computation}D. Gottesman, I. L. Chuang,\emph{Demonstrating the viability of universal quantum computation using teleportation and single-qubit operations}, Nature {\bf402}, 390–393 (1999).

\bibitem{Boschi98}D. Boschi, S. Branca, F. De Martini, L. Hardy, and S. Popescu, \emph{Experimental Realization of Teleporting an Unknown Pure Quantum State via Dual Classical and Einstein-Podolsky-Rosen Channels}
Phys. Rev. Lett. {\bf80}, 1121 (1998).

\bibitem{Bownmeester97}D. Bouwmeester, J.-W. Pan, K. Mattle, M. Eibl, H. Weinfurter, and A. Zeilinger, \emph{Experimental quantum teleportation}, Nature {\bf390}, 575-579 (1997).

\bibitem{Furusawa98}A. Furusawa, et al. \emph{Unconditional quantum teleportation}, Science {\bf282}, 706–709 (1998).

\bibitem{TelepReview}S. Pirandola, J. Eisert, C. Weedbrook, A. Furusawa, and S. L. Braunstein, \emph{Advances in quantum teleportation}, Nature Photonics {\bf9}, 641–652 (2015).


\bibitem{NielsenChuang}M. A. Nielsen and I. L. Chuang, \emph{Quantum Computation
and Quantum Information} (Cambridge University Press, 2000).

\bibitem{Horodecki99}M. Horodecki, P. Horodecki, and R. Horodecki, \emph{General teleportation channel, singlet fraction, and quasidistillation}, Phys. Rev. A {\bf60}, 1888 (1999).

\bibitem{Masanes}Ll. Masanes, \emph{All Bipartite Entangled States Are Useful for Information Processing}, Phys. Rev. Lett. 96, 150501 (2006).

\bibitem{footnote1} Note that we take the data $\{\sigma_{a|\omega_x}^\rB\}_{a,x}$ to implicitly contain as data the states to be teleported $\{\omega_x^\rV\}_x$.

\bibitem{footnote2} One can also define other notions of teleportation robustness in a straightforward manner, by changing the white noise to more general types of noise. We focus here on the random robustness only for ease of presentation.
    
\bibitem{footnote 3} A matlab code implementing this SDP is available at https://github.com/paulskrzypczyk/nonclassicalteleportation.

\bibitem{EntReview} R. Horodecki, P. Horodecki, M. Horodecki, and K. Horodecki, \emph{Quantum entanglement}, Rev. Mod. Phys. {\bf81}, 865 (2009).

\bibitem{Peres96}A. Peres, \emph{Separability Criterion for Density Matrices}, Phys. Rev. Lett. {\bf77}, 1413 (1996).

\bibitem{VB94}S. Boyd and L. Vandenberghe, \emph{Convex Optimization}, (Cambridge University Press, 2004).

\bibitem{cvx} M. Grant, S. Boyd, \textit{CVX: Matlab Software for Disci-
plined Convex Programming}, version 2.1, \url{http://cvxr.
com/cvx} (2014).

\bibitem{Horodecki96}M. Horodecki, P. Horodecki, R. Horodecki, \emph{Separability of Mixed States: Necessary and Sufficient Conditions}, Physics Letters A {\bf223}, 1 (1996).

\bibitem{DPS02}A. C. Doherty, Pablo A. Parrilo, and F. M. Spedalieri, \emph{Distinguishing Separable and Entangled States}, Phys. Rev. Lett. {\bf88}, 187904 (2002).

\bibitem{supp}See Supplemental Material.

\bibitem{UPB} C. H. Bennet, D. P. DiVincenzo, T. Mor, P. W. Shor, J. A Smolin, B. Terhal, \emph{Unextendible product bases and bound entanglement},  Phys. Rev. Lett. {\bf82}, 5385 (1999).


\bibitem{Guhne} O. G\"{u}hne , P. Hyllus , D. Bruss , A. Ekert , M. Lewenstein , C. Macchiavello , A.
Sanpera, \emph{Detection of entanglement with few local measurements}, Phys. Rev. A \textbf{66}, 062305 (2002).

\bibitem{Robustness}Guifre Vidal, Rolf Tarrach, \emph{Robustness of entanglement}, Phys.Rev. A {\bf59} 141 (1999).

\bibitem{LCS99} N. L\"{u}tkenhaus, J. Calsamiglia, and K.-A. Suominen, \emph{Bell measurements for teleportation}, Phys. Rev. A {\bf59}, 3295 (1999).


\bibitem{footnote4} Notice that the denominator of the RHS is non-negative.  This is because for a general set of input states $\overline{F}_\mathrm{cl}
\geq 1/d$,  as $1/d$ is achieved by the trivial strategy whereby Bob sends $\rho_{a|\omega_x}^\rB = \openone^\rB/d$ to the verifier, and Alice outputs with any probability distribution $p(a|\omega_x)$.

\bibitem{Grudka} J. Modlawska and A. Grudka, \emph{Increasing singlet fraction with entanglement swapping}, Phys. Rev. A 78, 032321 (2008).

\bibitem{MassarPopescu95}S. Massar and S. Popescu, \emph{Optimal Extraction of Information from Finite Quantum Ensembles}, Phys. Rev. Lett. {\bf74}, 1259 (1995).


\bibitem{Steering} H.~M.~Wiseman, S.~J.~Jones, and A.~C.~Doherty, \textit{Steering Entanglement, Nonlocality, and the Einstein-Podolsky-Rosen Paradox}, Phys. Rev. Lett. \textbf{98} 140402 (2007).

\bibitem{Steering Review}D. Cavalcanti and P. Skrzypczyk, \emph{Quantum steering: a short review with focus on semidefinite programming,  Rep. Prog. Phys. 80 024001 (2017).}

\bibitem{Buscemi} F. Buscemi,\emph{ All Entangled Quantum States Are Nonlocal}, Phys. Rev. Lett. {\bf108}, 200401 (2012).


\bibitem{MT}M. T. Quintino, T. V\'ertesi, D. Cavalcanti, R. Augusiak, M. Demianowicz, A. Ac\'in, N. Brunner, \emph{Inequivalence of entanglement, steering, and Bell nonlocality for general measurements}, Phys. Rev. A 92, 032107 (2015).

\bibitem{CHW13}Eric G. Cavalcanti, Michael J. W. Hall, Howard M. Wiseman, \emph{Entanglement verification and steering when Alice and Bob cannot be trusted}, Phys. Rev. A 87, 032306 (2013)-

\bibitem{MDI-EW}C. Branciard, D. Rosset, Y.-C. Liang, N. Gisin, \emph{Measurement-Device-Independent Entanglement Witnesses for All Entangled Quantum States}, Phys. Rev. Lett. {\bf110}, 060405 (2013).


\bibitem{footnote3} Notice that, although the definition of $\tilde{\sigma}^{\rV\rB}$ appears to depends on $a$,  it is in fact independent of $a$, which follows from the left-hand-side of \eqref{e: Mtilde}).

\bibitem{UPB} C. H. Bennet, D. P. DiVincenzo, T. Mor, P. W. Shor, J. A Smolin, B. Terhal, \emph{Unextendible product bases and bound entanglement},  Phys. Rev. Lett. {\bf82}, 5385 (1999).

\bibitem{Horodecki99}M. Horodecki, P. Horodecki, and R. Horodecki, \emph{General teleportation channel, singlet fraction, and quasidistillation}, Phys. Rev. A {\bf60}, 1888 (1999).


\bibitem{Guhne} O. G\"{u}hne , P. Hyllus , D. Bruss , A. Ekert , M. Lewenstein , C. Macchiavello , A.
Sanpera, \emph{Detection of entanglement with few local measurements}, Phys. Rev. A \textbf{66}, 062305 (2002).


\end{thebibliography}
\end{document}